\documentclass[a4paper]{jpconf}
\usepackage{graphicx}
\usepackage{xspace}
\usepackage{amsfonts,amsmath,stackrel}

\renewcommand{\br}[1]{\left(#1\right)}
\newcommand{\ud}[1]{\mathrm{d}#1}

\newcommand{\sq}[1]{\left[#1\right]}

\newcommand{\abs}[1]{\left|#1\right|}

\renewcommand{\vec}[1]{\boldsymbol{#1}}

\newcommand{\qq}{\mathcal{Q}}

\newcommand{\rr}{\mathcal{R}}

\newcommand{\dx}{\dot{x}}
\newcommand{\da}{\dot{a}}
\newcommand{\db}{\dot{b}}
\newcommand{\dk}{\dot{k}}
\newcommand{\veri}{\mathbf{e}_1}
\newcommand{\verj}{\mathbf{e}_2}
\newcommand{\vern}{\mathbf{n}}
\newcommand{\dveri}{\dot{\mathbf{e}}_1}
\newcommand{\dverj}{\dot{\mathbf{e}}_2}
\newcommand{\dvern}{\dot{\mathbf{n}}}
\newcommand{\inline}[1]{\begin{center}\mbox{${#1}$}\end{center}}
\newcommand{\elm}{m}
\newcommand{\ux}{u_{,\,x}}
\newcommand{\uy}{u_{,\,y}}
\newcommand{\uxx}{u_{,\,xx}}
\newcommand{\uxy}{u_{,\,xy}}
\newcommand{\uyy}{u_{,\,yy}}
\newcommand{\oo}{\omega}
\newcommand{\ox}{\omega_{,\,\xi}}

\newcommand{\oxx}{\omega_{,\,\xi\xi}}
\newcommand{\oxy}{\omega_{,\,\xi\eta}}
\newcommand{\oyy}{\omega_{,\,\eta\eta}}
 \newcommand{\spliteq}[1]{\begin{align}\begin{split} #1
       \end{split}
      \end{align}}

\begin{document}
\title{Spinor particle. An indeterminacy in the motion
of relativistic dynamical systems with separately fixed mass and spin.\footnote{This article is based on my talk \textit{Spinor particle --
Unexplained abnormality in the motion
of Fundamental Dynamical Systems} at the Seventh International Conference \textit{Quantum Theory and Symmetries} (QTS-7), August 2011, Prague. It includes also some new material.}}

\author{{\L}ukasz Bratek}

\address{\textit{The H. Niewodnicza{\'n}ski Institute of Nuclear Physics,
Polish Academy of Sciences, Radzikowskego {152}, {31-342} Krak{\'o}w, Poland}}

\ead{lukasz.bratek@ifj.edu.pl \hspace{2cm} \textbf{IFJPAN-IV-2011-10}}

\begin{abstract}
We give an argument that a broad class of geometric models of spinning relativistic particles with Casimir mass and spin being separately fixed parameters, have indeterminate worldline (while other spinning particles have definite worldline). This paradox suggests that for a consistent description of spinning particles something more general than a worldline concept should be used.  As a particular case, we study at the Lagrangian level the Cauchy problem for a spinor particle  and then, at the constrained Hamiltonian level, we generalize our result to other particles.

\end{abstract}

\section{Introduction}
The purpose of the present work is to show that a class of geometric models of spinning relativistic particles with Casimir mass and Casimir spin being separately fixed parameters of these models, have indeterminate free motion, at least when the rapidity, defined as the hyperbolic angle between the 4-velocity and the canonical 4-momentum, is finite. We
do not address the quite separate and critical case of spinning particles moving with the speed of light and still having nonzero mass, that is, classical analogs of zitterbewegung states of Dirac electron.


   Our analysis concerns point particles with intrinsic structure described by an arbitrary number of 4-vectors.
  This general scheme can encompass a broad class of classical models of relativistic
particles with spin (a brief review of particular models of relativistic particles with spin
can be found in \cite{bib:Frydryszak}). Our scheme includes also spinorial particles described by a system of spinors since each spinor can be described equivalently by a triad of four-vectors satisfying some conditions. In particular, we propose a particle model with internal structure described by a single spinor of fixed magnitude and with spinor phase being a cyclic variable. Its physical state is described by 6 physical degrees of freedom, 3 for position in space and 3 Euler angles describing the spinor direction and the spinor phase.  Then we study the Cauchy problem for this particle and come to the conclusion that the Hessian rank is lower than $6$ when the condition of separately fixed mass and spin is imposed. Finally, at the constrained Hamiltonian level we give an argument that the Hessian rank will be reduced also for other vectorial particles with separately fixed mass and spin.

 Reparametrization invariance (a requirement for all relativistic systems) implies there is an arbitrary function present in the solutions, but this arbitrariness does not affect the worldline as a path in spacetime. However, when the Casimir mass and spin are separately fixed, independently of the initial conditions, an additional arbitrary function can appear. In effect, the motion is not unique (the equations of motion are under-determined) despite the worldline parameter has been fixed. One has to admit that this is a defect of a dynamical system since its trajectory in spacetime should be unique to the extent of choosing parametrization.   It seems doubtful that Lagrangians of such particles could be reinterpreted as particles with a lower number of physical degrees of freedom and determinate worldline.

 As an aside remark, note that there is another interpretation possible. We see there is a one-fold family of distinct worldlines
   possible for the same physical state of a particle with separately fixed mass and spin. For example, in the case of the simplest kind of such a particle, consisting of a worldline and a single null direction, the union of all worldlines evolving from an initial state forms a single tube in spacetime (see the example at the end of this paper).
    The two arbitrary functions have the nature of gauge variables, which is consistent with the arbitrariness  of choosing coordinates on that tube. In particular, these coordinates can be chosen so as they are conjugate to the Casimir mass and spin, respectively. 
   It would be interesting for the future to reinterpret particles with spin as two dimensional objects in spacetime, since then a unique correspondence of such objects and physical states could be regained. This would require to write action functional as a two-fold integral over a tube-like object in spacetime, not a line integral over a world-line. A combination of parameters of spin and mass gives rise to a fundamental length scale. There is no room for such a scale on a worldline, whereas the scale is natural for a tube. The inconsistency of motion we encounter for dynamical systems with separately fixed mass and spin (and resulting paradoxes) might be a signal that we have inappropriate model of spinning particles.

\section{A class of models of particles with spin}

The 4-momentum $p$ of a dynamical system arises in the canonical formalism by means of infinitesimal translations in spacetime,  $p_{\mu}=-\partial_{\dot{x}^{\mu}}\mathcal{L}$ with $\mathcal{L}$ being the Lagrangian and $\dot{x}^{\mu}$ the tangent vector to the worldline (we use the minus sign convention due to the assumed signature of the metric tensor [1,-1,-1,-1]).
 The direction of $p^{\mu}$ defines the time axis of the center of momentum frame.  The first Casimir scalar $p_{\mu}p^{\mu}$ defines the invariant mass of the system, called Casimir mass.
The intrinsic spin  arises by means of infinitesimal rotations of the system in
the center of momentum frame.
 In the canonical formalism the spin is described with the help of an
 antisymmetric tensor of intrinsic spin $J_{\mu\nu}$. It is obtained by projecting the angular momentum  $M_{\mu\nu}$ onto the subspace orthogonal to $p$ $$J_{\mu\nu}=M_{\alpha\beta}h^{\alpha}_{\phantom{\alpha}\mu}h^{\beta}_{\phantom{\beta}\nu},\qquad h^{\mu}_{\phantom{\mu}\mu}=\delta^{\mu}_{\phantom{\mu}\mu}-\frac{p^{\mu}p_{\nu}}{p^2}.$$
 Owing to its algebraic properties, $J_{\mu\nu}=-J_{\nu\mu}$ and $J_{\mu\nu}p^{\nu}=0$, tensor $J_{\mu\nu}$
  can be equivalently described in terms of the Pauli-Luba\'{n}ski spin-pseudovector
  $W^{\mu}=-\frac{1}{2}\epsilon^{\mu\alpha\beta\gamma}M_{\alpha\beta}p_{\gamma}$.
  Then, the second Casimir invariant can be expressed as  $$W^{\mu}W_{\mu}=-\frac{1}{2}p^{\alpha}p_{\alpha}J^{\mu\nu}J_{\mu\nu}.$$
For the spin to be nonzero, an additional structure  must be associated with the worldline of a dynamical system.
Let this structure be encoded in terms of several additional vectors $q_i$, $i=1,2,\dots,N_v$. This generalizes Barut's idea of modeling particles with spin \cite{bib:barut}.
Written in terms of canonical momenta $\stackrel[i]{}{\pi}{}_{\!\!\mu}=-\partial_{\stackrel[i]{}{\dot{q}}{}_{\mu}}\mathcal{L}$,  the tensor of spin attains the form
 \newcommand{\qa}{\stackrel[i]{}{q}{}_{\!\!\mu}}
\newcommand{\qb}{\stackrel[i]{}{q}{}_{\!\!\nu}}
\newcommand{\pia}{\stackrel[i]{}{\pi}{}_{\!\!\mu}}
\newcommand{\pib}{\stackrel[i]{}{\pi}{}_{\!\!\nu}}
\newcommand{\pa}{{p}_{\mu}}
\newcommand{\pb}{{p}_{\nu}}
\newcommand{\pq}{p\!\stackrel[i]{}{q}}
\newcommand{\ppi}{p\!\stackrel[i]{}{\pi}}
  $$p^2 J_{\mu\nu}=\sum\limits_{i=1}^{N_v}\sq{p^2\br{\qa\pib-\pia\qb}+
  {\ppi}\br{\pa\qb-\qa\pb}+{\pq}\br{\pia\pb-\pa\pib}}.$$
It can be verified by a direct calculation that the  spin invariant is
$$W^{\mu}W_{\mu}=-\sum_{i,j=1}^{N_v}\det\left[\begin{array}{ccc}
pp&p\stackrel[j]{}{q}&p\stackrel[j]{}{\pi}\\
\stackrel[i]{}{q}p&\stackrel[i]{}{q}\stackrel[j]{}{q}&\stackrel[i]{}{q}\stackrel[j]{}{\pi}\\
\stackrel[i]{}{\pi}p&\stackrel[i]{}{\pi}\stackrel[j]{}{q}&\stackrel[i]{}{\pi}\stackrel[j]{}{\pi}
\end{array}\right].$$
There is variety of particle models that can be described by a set of vectors.
As a particular realization one can mention KLS spinning particle \cite{bib:segal} equivalent to fundamental relativistic rotator \cite{bib:astar1}  and generalizations  \cite{bib:segal2}\cite{bib:bratek2}, or rotating tetrad of Hanson and Regge \cite{bib:hanson} which can be equivalently viewed   as a dynamical system consisting of a worldline and associated with it three null directions.

 In this paper, we will be studying another
 example of particles with spin.
Their intrinsic structure is described by
a triad of mutually orthogonal vectors of which one is null
    and the other two are spacelike and unit. Such a triad describes a single spinor. Therefore, we call such models \textit{spinor particles}. They can have at most $7$ physical degrees of freedom, 3 for the position of the center of mass and 4 additional degrees of freedom pertinent to a spinor: spinor direction (determined by two spherical angles), spinor magnitude and spinor phase. We will focus our attention on a spinor particle with $6$ degrees of freedom with spinor magnitude regarded as a gauge variable, irrelevant to the particle's structure, which, for reasons explained later, we call \textit{relativistic axisymmetric top}. Complementary to it is a family of particle models studied already in a similar manner in \cite{bib:bratek2} with irrelevant spinor phase.

 To reduce the variety of Lagrangians a priori possible for geometric models of particles and to fix a unique action functional,  it is postulated as a basic principle  that both Casimir
invariants of the Poincar\'{e} group should be fixed parameters. This idea was originally used in \cite{bib:segal} and called \textit{strong conservation} and later was independently arrived at in \cite{bib:astar1} based on the Wigner
idea of classification of relativistic quantum-mechanical systems \cite{bib:wigner}. Staruszkiewicz pointed
out \cite{bib:astar1} that quantum irreducibility, being a simple algebraic notion, has its classical counterpart
(unlike unitarity) which means in the context of relativistic classical mechanical systems
that both Casimir invariants of the Poincar\'{e} group should be parameters with fixed numerical
values (that is, be independent of the state of motion like for a quantum particle).
 It is convenient to chose the two parameters so as \cite{bib:astar1}
\begin{equation}\label{eq:qts7_FundCond}p_{\mu}p^{\mu}=m^2,\qquad  W_{\mu}W^{\mu}=-\frac{1}{4}m^4\ell^2.\end{equation}
These two conditions for the action integral are referred to as  \textit{fundamental conditions}. It turned out that the motion of dynamical systems with this particular property is not completely determined by the stationary action principle and the initial conditions \cite{bib:bratek3}. There is a premise that this non-uniqueness can be general.
Later, we shall shed some light on this issue at the Hamiltonian level.

  \section{Spinor particles and axisymmetric relativistic top}
\begin{figure}[h]
\includegraphics[width=0.32\textwidth]{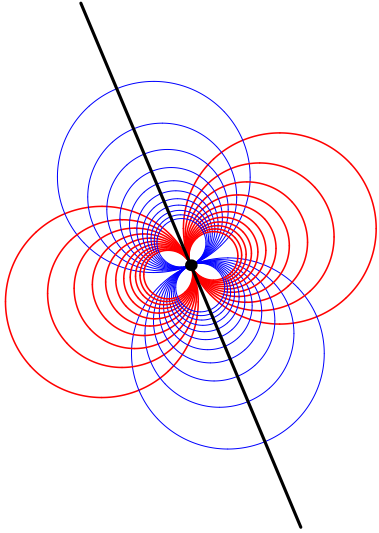}\hspace{2pc}
\begin{minipage}[b]{14pc}\caption{\label{fig:spinor} Two orthogonal parabolic pencils of circles crossing at a single point shown on the sphere of complex numbers. The circles are images of two families of spatial directions $a+\lambda_ak$ and $b+\lambda_bk$ enumerated by real numbers $\lambda_a,\lambda_b$.
The crossing point is the image of null direction $k$. Its position represents the direction of spinor in space. Rotations around the image of $k$ represents changes in the spinor phase. A single parabolic bunch of circles is the graphical representation of the notion of spinor flag introduced by Penrose \cite{bib:penrose}.
}
\end{minipage}
\end{figure}

Crucial to our construction of a  spinor particle is
the possibility of describing a single spinor by means of a  triad of vectors in Minkowski space.
 A skew-symmetric tensor  $F_{\mu\nu}$ (bivector) is called null when $F_{\mu\nu}F^{\mu\nu}=0$ and $\epsilon^{\alpha\beta\mu\nu}F_{\alpha\beta}F_{\mu\nu}=0$.
The following Cartan-Whittaker relation \cite{bib:cartan}\cite{bib:whittaker} establishes a $2:1$ correspondence between spinors $u^{A}$ and null bivectors $F^{\mu\nu}$
$$iF^{01}-F^{23}=\br{u^0}^2-\br{u^1}^2,\qquad iF^{02}-F^{31}=i\sq{\br{u^0}^2-\br{u^1}^2},
\qquad iF^{03}-F^{12}=-2u^0u^1.$$
The Cartan-Whittaker tensor can be regarded as a spatiotemporal interpretation of a spinor. On account that both fundamental invariants of tensor $F_{\mu\nu}$ vanish, there is a null vector $k$ and two unit spatial vectors such that \cite{bib:astar2}
\begin{eqnarray*}
&F^{\mu\nu}=k^{\mu}a^{\nu}-k^{\nu}a^{\mu}=\epsilon^{\mu\nu\alpha\beta}k_{\alpha}b_{\beta},&\\
&kk=ak=bk=0,\qquad ab=0,\qquad aa=bb=-1.
\end{eqnarray*}Then $k^{\mu}=u^{+}\sigma^{\mu}u$, $(k^0>0)$. Vectors $a,b$ are not unique. There is a whole family consisting of vectors of the form $$a+\lambda_a k,  \qquad b+\lambda_b k,\qquad -\infty<\lambda_a,\lambda_b<+\infty.$$ Using this correspondence, a spinor can be visualized on the sphere of complex numbers by two pencils of parabolic circles crossing at a single point, see Fig.\ref{fig:spinor}.

\medskip

We wish to obtain a dynamical system akin to a rotating rigid body with 6 degrees of freedom, a rotating rigid triad. Geometrically, we want it to be the simplest generalization of a relativistic rotator \cite{bib:astar1} and simpler in structure than the spherical top of Hanson and Regge \cite{bib:hanson}. As so, it is best we choose the spinor magnitude as irrelevant in our construction. Then, there must be a projection symmetry $k\to \lambda k$, with arbitrary function $\lambda$.
Thus, the particle's structure is not altered by performing
 the transformations \inline{k\to\lambda k,\quad a\to a+\mu k,\quad b\to b+\nu k,} with arbitrary functions $\lambda$, $\mu$ and $\nu$. These three transformations must be gauge symmetries of the action integral. In that case  $\lambda$, $\mu$ and $\nu$ are gauge variables.
  When  the spinor phase is increased by an angle $\psi/2$, vectors $a$ and $b$ get rotated through angle $\psi$ about $k$ in the plane spanned by the original vectors $a$ and $b$: $$a\to a\cos{\psi}+b\sin{\psi},\qquad b\to -a\sin{\psi}+b\cos{\psi},\qquad k\to k.$$ We see, that the particle has three intrinsic physical  degrees of freedom. They can be parameterized by three Euler angles. Two spherical angles $\theta,\phi$ describing the direction of the spinor and the third angle, $\psi$, describing the spinor phase.

In what follows we will construct Hamilton's action for our particle.
Relativistic invariance of the action ensures 10 integrals of motion, the components of
$p^{\mu}$ and $M_{\mu\nu}=-M_{\nu\mu}$. In order to have this invariance, the action should be built up out of reparametrization invariant Lorentz scalars not involving variables $x^{\mu}$. This is a general requirement for a relativistic particle model.
Further, for dimensional reasons
the minimum number of free parameters is two (we work with units in which $c=1$).
We can choose a unit of mass $\elm$ and a unit of length $\ell$ as such parameters. We assume these two are the only free parameters of our particle model. In addition, we assume that the spinor phase should be a cyclic variable. That means that the spinor phase cannot be  present in the Lagrangian explicitly. With these simple and clear requirements the variety of possible Lagrangians for our  particle has been reduced to a minimum.

To satisfy the above requirements, the action integral for the spinor particle must be constructed out of reparametrization and gauge invariant Lorentz scalars of appropriate dimension. These scalars must be some functions of several basic scalars, the latter being scalar products of the worldline tangent $\dot{x}^{\mu}$, of the three vectors defining the particle's intrinsic structure $a^{\mu}$, $b^{\mu}$ and $k^{\mu}$, and of their derivatives $\dot{a}^{\mu}$, $\dot{b}^{\mu}$ and $\dot{k}^{\mu}$ with respect to any worldline parameter.
Since the Gram determinant of vectors $a,b,k,\dx$, which equals to  $-(k\dx)^2$, is always nonzero, vectors $a,b,k,\dx$ form a basis in Minkowski space. The velocities $\da$, $\db$ and $\dk$ can be decomposed in this basis. Thus, there is at most $10$ nonzero scalar products that are independent:
 $k\da$, $ k\db$, $ a\db$, $ a\dx$, $ b\dx$, $ k\dx$, $ \da\dx$, $ \db\dx$, $ \dk\dx$, $ \dx\dx$.
  The other scalars are their functions, for example $$\dk\dk=-(k\da)^2-(k\db)^2,\qquad \da\dk=-a\db\,k\db+\frac{k\da\,\dk\dx}{k\dx}-\frac{a\dx\,(k\da)^2}{k\dx}-
 \frac{b\dx\,k\da\,k\db}{k\dx},$$ or are some constraints resulting from the assumptions made for $a$, $b$ and $k$.
On gauge transforming, any Lorentz scalar becomes a function of six variables $\mu$, $\nu$, $\lambda$, $\dot{\mu}$, $\dot{\nu}$ and $\dot{\lambda}$. We do not want these functions to appear in the action. This gives us $6$ conditions for a gauge invariant function $\chi$: $\partial_{\lambda}\chi=0$, $\partial_{\dot{\lambda}}\chi=0$, and similarly for $\mu$ and $\nu$. As so, $\chi$ will be a function of only four independent basic gauge invariants. Three of them can be easily guessed: the worldline element $I_0=\dx\dx$, and two velocities $I_1=\frac{k\da}{k\dx}$ and $I_2=\frac{k\db}{k\dx}$ associated with rotations about $b$ and $a$, respectively.
 Then, we expect the fourth invariant $I_3$ to be associated with a rotation about direction $k$.
  Therefore, it should contain the scalar $a\db=-\dot{a}b$. But it is not gauge invariant, it transforms as $a\db\to a\db+\mu\, I_2\, k\dx-\nu\, I_1\, k\dx$. The compensating counter-terms needed to remove this gauge dependence must have the same physical dimension. These are $I_1\,b\dx$ transforming to $I_1\,b\dx+\nu\,I_1\,k\dx$ and $I_2\,a\dx$ transforming to $I_2\,a\dx+\mu\, I_2\,k\dx$. Hence, $I_3\equiv{}a\db-I_2\, a\dx+I_1\, b\dx$ is gauge invariant. Furthermore, when $a$ and $b$ get rotated through an angle $\psi$ then $I_1\to I_1\,\cos{\psi}+I_2\,\sin{\psi}$ and $I_2\to-I_1\,\sin{\psi}+I_2\,\cos{\psi}$.
 The scalars  $I_1$ and $I_2$ are not explicitly phase-invariant, however, $I_1^2+I_2^2=-\dk\dk$ is. Under the same rotation $I_3$ transforms as $I_3\to I_3+\dot{\psi}$, in accord with the expectation that $I_3$ describes the velocity associated with the phase. It is also clear from this transformation law that $I_3$ does not explicitly depend on the phase.
This way we obtain two dimensionless, reparametrization- and gauge-invariant scalars
 $$\rr={\frac{\ell}{\sqrt{\dx\dx}}\frac{(a\db)(k\dx)+(k\da)(b\dx)+(b\dk)(a\dx)}{
 k\dx}},\quad \qq=\ell^2\frac{(a\dk)^2+(b\dk)^2}{(k\dx)^2}\equiv-\ell^2\frac{\dk\dk}{(k\dx)^2}.$$
The most general scalar formed out of $I_0$, $\rr$ and $\qq$ which reduces to a worldline element when $\ell=0$ is $\sqrt{I_0}\,F(\rr,\qq)$ with an arbitrary and sufficiently smooth function $F$.
This way we arrive at the following most general action functional satisfying all our requirements
 \begin{equation}\label{eq:HamiltonsAction}
 S=-\elm\int\ud{\tau}\sqrt{\dx\dx}\,
F(\mathcal{P},\mathcal{Q}).\end{equation}We call  this model of a spinor particle a \textit{relativistic axisymmetric top} on account the distinct role played by null direction $k$ and the pair
of spatial directions $a$ and $b$ and the rotational symmetry about an axis determined by $k$.

 We can gain better insight into the physical meaning of invariants $\rr$ and $\qq$ in terms of angular velocity vector $$\Omega=\frac{(a\db)\,k+(k\da)\,b+(b\dk)\,a}{k\dx}.$$ Then
   $\rr=\ell\,{\Omega \frac{\dot{x}}{\sqrt{\dot{x}\dot{x}}}}$ and $\qq=-\ell^2\,\Omega\Omega$.  In a particular gauge in which \begin{equation}\label{eq:Gauge}\dot{x}=[1,\vec{v}],\quad a=[0,\veri],\quad b=[0,\verj],\quad  k=[1,\vern],\end{equation} with  $\veri$, $\verj$ and $\vern\equiv\veri\times\verj$ forming an orthonormal rigid frame,
  $$\rr={\frac{\ell}{\sqrt{1-\vec{v}^2}}\cdot{\frac{\vec{\omega}(\vern-\vec{v})}{
 1-\vern\vec{v}}}},\qquad \qq=\ell^2\frac{\vec{\omega}^2-(\vec{n}\vec{\omega})^2}{\br{1-\vern\vec{v}}^2}.$$
 Here, vector
  $\vec{\omega}\equiv\veri(\dverj\vern)+\verj(\dvern\veri)
 +\vern(\dveri\verj)$
  is the angular velocity of rotation of the rigid frame.
  In terms of Euler angles,  $\vec{n}=[\sin{\theta}\cos{\phi},\sin{\theta}\sin{\phi},\cos{\theta}]$,
  $\dverj\vern=\dot{\theta}\cos{\psi}+\dot{\phi}\sin{\theta}\sin{\psi}$,
 $\dvern\veri=-\dot{\theta}\sin{\psi}+\dot{\phi}\sin{\theta}\cos{\psi}$ and
 $\omega_{||}=\dveri\verj=\dot{\psi}+\dot{\phi}\cos{\theta}$.
 In the  limit $\abs{\vec{v}}\ll1$, $\rr^2/\ell^2\approx\omega_{||}^2$ and $\qq/\ell^2\approx\omega_{\bot}^2$. Thus,  $\rr$ and $\qq$ measure the angular rotation
 in a Lorentz covariant way. The scalar $\qq/\ell^2$ refers to the angular velocity of
   the image of $k$ on a unit sphere centered at some point moving with velocity $\dot{x}$, while the scalar $\rr/\ell$ refers to the angular velocity of rotation about an instantaneous axis of rotation determined by $k$.

 \section{The Cauchy problem and fundamental conditions}
 To make the notation more transparent, we introduce a function $u$ such that
$$F(\qq,\rr)=u\br{x,y},\qquad x=\sqrt{\qq}>0,\quad y=\rr.$$ The
  Casimir invariants for Hamilton's action \eqref{eq:HamiltonsAction} read (\textit{we assume that $\dot{x}\dot{x}>0$})
  \spliteq{
 C_M&=\frac{pp}{\elm^2}=\br{u-y\,\uy-x\,\ux}^2-x^2\br{\ux^2+\uy^2}\\
     C_J&=\frac{WW}{-\frac{1}{4}\elm^4\ell^2}=4\br{u-y\,\uy}^2\br{\ux^2+\uy^2}
         } In general, the invariants will be some complicated functions of the initial conditions: both $C_M$ and $C_J$ will depend on the physical state.

     From now on we will be assuming that the action integral has been expressed in terms of only the coordinates referring
to the  degrees of freedom considered physical. For the spinor particle it is convenient to work in the gauge \eqref{eq:Gauge}.  In that gauge the worldline parameter is identified with $\tau=t\equiv x^0$ and the physical state is determined by specifying positions $q=\{x^1,x^2,x^3,\theta,\phi,\psi\}$ and the corresponding velocities. This way all arbitrary degrees of freedom can be removed, and the covariant action integral \eqref{eq:HamiltonsAction} reduced to the ordinary non-covariant form with a reduced Lagrangian $L$ $$S=\int L(q,\dot{q})\,dt.$$
  For a physical state to be unique it is essential that the reduced Lagrangian be regular. That means that the determinant of the Hessian matrix assigned to $L$ $$\mathcal{H}=\sq{\frac{\partial^2L}{\partial{\dot{q}^i}\partial{\dot{q}^j}}}$$ should be nonzero.
          By a cumbersome calculation similar to that presented in detail in \cite{bib:bratek2}, one can show that (to the extent of an unimportant nonzero kinematical factor independent of $u$ and dependent on the particular map) the Hessian determinant is
 $$\det{\mathcal{H}}=\frac{\br{u-y\,\uy}\br{x\,\uy^2+\br{u-y\,\uy}\ux}}{x}
\br{\br{\ux^2+\uy^2}\uyy+\br{u-y\,\uy}\br{{\uxx\uyy-\uxy^2}}}.$$
         By a direct calculation, one can show an interesting relationship between the Hessian
determinant and a Jacobian determinant of a mapping from coordinates $(x,y)$ to coordinates $(C_M,C_J)$
\spliteq{\det{\mathcal{H}}&=
\frac{\br{u-y\,\uy}\ux+x\,\uy^2}{16\,x^2\,\mathcal{E}_C}
\left|\frac{\partial\br{C_M,C_J}}{\partial\br{x,y}}\right|,\qquad \mathrm{when} \quad \mathcal{E}_C\ne0,\\ \mathcal{E}_C&\equiv \frac{u-y\,\uy}{x}\br{y\,\ux-x\,\uy}+\uy\,\br{x\,\ux+y\,\uy}.\nonumber}
 It is quite analogous to a similar relation found in \cite{bib:bratek2}. The Jacobian determinant is zero whenever the Casimir invariants are functionally dependent.
  In particular, this happens when fundamental conditions are imposed (we exclude the trivial example of a structureless particle with three degrees of freedom, in which case $C_J\equiv0$). We stress that the vanishing of the Jacobian does not necessarily imply the degeneracy of the Hessian, a counterexample was given in \cite{bib:bratek2}. But we can conclude, that the necessary condition for the Hessian could be nonzero when fundamental conditions \eqref{eq:qts7_FundCond} are imposed, is $\mathcal{E}_C=0$.

The task of solving the condition $\mathcal{E}_C=0$ simultaneously with fundamental conditions $C_M=1=C_J$ can be simplified considerably by means of the Legendre transformation for function $u$: $$u(x,y)\to x\,\xi+y\,\eta-\omega(\xi,\eta),\quad u_{,\,x}\to \xi,\quad u_{,\,y}\to\eta.$$ The Hessian determinant  is transformed to
$$\det{\mathcal{H}}=\frac{\oo-\xi\,\ox}{2\ox\br{\oxy^2-\oxx\,\oyy}}
\partial_{\xi}\br{\xi\,\oo-\br{\xi^2+\eta^2}\ox}^2.$$ This transformation is feasible for nondevelopable surfaces, that is, surfaces  for which ${\uxx\,\uyy-\uxy^2}\ne0$. The case of developable surfaces ${\uxx\,\uyy-\uxy^2}\equiv0$, in which case the inverse transformation from coordinates $(\xi,\eta)$ to $(x,y)$ is indefinite, must be treated separately.
\subsubsection*{$\uxx\,\uyy-\uxy^2\ne0$.} Equations $\mathcal{C}_M=1$ and $\mathcal{C}_J=1$ are transformed to $\oo^2-\br{\xi^2+\eta^2}\ox^2=1$ and $4\br{\xi^2+\eta^2}\br{\oo-\xi\,\ox}^2=1$, and can be solved algebraically for $\oo$ and $\ox$, giving
$$\omega\br{\xi,\eta}=\pm{\frac{\sqrt{\xi^2+\eta^2}+\epsilon\xi\sqrt{1-4\eta^2}}{2\eta^2}},\quad \epsilon^2=1.$$
Then, equation $\mathcal{E}_C=0$ is automatically satisfied.
Despite that, the Hessian determinant is still
vanishing, since for this solution $\br{\xi\,\oo-\br{\xi^2+\eta^2}\ox}^2={1-4\,\eta^2}$, which is independent of $\xi$.
\subsubsection*{$\uxx\,\uyy-\uxy^2=0$.}
Equations $\partial_yC_{M}=0$ and $\partial_yC_{J}=0$ are homogenous and linear in $\uxy$ and $\uyy$.
The determinant of this system is $-16x\br{u-y\,\uy}^2\mathcal{E}_C$, and it vanishes when $\mathcal{E}_C=0$ ($u-y\,\uy$ must be nonzero when $C_J=1$).
When $\mathcal{E}_C\ne0$, we  have $\uxy=0$ and $\uyy=0$. Then $u(x,y)=\tilde{\kappa} y+f(x)$ with $\tilde{\kappa}$ being a constant. Solution of fundamental conditions is now easy, and we obtain $u=\pm\sqrt{1-\frac{x^2}{4}\sin^2{\kappa}+x \cos{\kappa}}+\frac{y}{2}\sin{\kappa}$, where we specified $\tilde{\kappa}$. Unfortunately, for $\mathcal{E}_C\ne0$ the Hessian determinant must vanish (in particular, for $\sin{\kappa}=0$ we obtain fundamental relativistic rotator \cite{bib:astar1} or KLS particle \cite{bib:segal} in which case the Hessian rank is $4$ \cite{bib:bratek3}).
Further, it can be demonstrated that there is no solution for which $\mathcal{E}_C=0$. To this end
we  employ the general solution of equation  $\uxx\,\uyy-\uxy^2=0$. Geometrically, every such solution is a developable surface of the form  $u=\alpha\, x+w(\alpha)\, y+ v(\alpha)$ \cite{bib:hilbert}. Functions  $v$ and $w$ are arbitrary, whereas function $\alpha$ is implicitly defined by condition $0=x+w'(\alpha)\, y+v'(\alpha)$, or $\alpha$ is constant.
Partial differentiation of condition $\mathcal{C}_M=1$ with respect to $x$ and $y$ gives
  $-2\,x\br{\alpha^2+w^2}-2\,\beta\,\alpha_{,\,x}=0$ and $-2\,{\beta}\,\alpha_{,\,y}=0$, where $\beta\equiv v\br{x+y\, w'(\alpha)}+x^2\br{\alpha+w\,w'(\alpha)}$. Suppose that $\alpha_{,\,y}\ne0$, then the second equation gives $\beta=0$ and, accordingly, the first equation will be satisfied only when $\alpha=0$ and $w=0$, a contradiction with $\alpha_{,\,y}\ne0$. Thus, $\alpha_{,\,y}=0$, that is, $\alpha$ is a function of $x$ only,  $\alpha=\gamma(x)$. By taking into account that $u_{,\,x}=\alpha$ and $u_{,\,y}=w(\alpha)$, and that for $\mathcal{C}^2$ solutions  $u_{,\,xy}=u_{,\,yx}$, we obtain $w'(\gamma)\,\gamma'(x)=0$. Function $\gamma$ cannot be constant, since the plane $u=a\, x+b\, y+c$ cannot solve fundamental conditions. Hence, $w'(\gamma)=0$ and therefore $u=x\, \gamma(x)+b\, y+v(\gamma(x))$ with a condition $x+v'(\gamma)=0$. But this solution is of the form  $u=b\, y+g(x)$ we have already investigated, for which $\mathcal{E}_C\ne0$.

\medskip
Above, we have shown that there are two essentially different Lagrangians for axisymmetric top \eqref{eq:HamiltonsAction} satisfying fundamental conditions \eqref{eq:qts7_FundCond}. However, the vanishing of the Hessian determinant in both cases shows the systems are defective as dynamical systems with $6$ physical degrees of freedom.

\section{Fundamental dynamical systems and constrained Hamiltonians}

At the Lagrangian level there is no obvious reason for why fundamental conditions \eqref{eq:qts7_FundCond} can imply Hessian singularity and the calculation leading to such conclusion is quite involved. The situation is more clear at the Hamiltonian level.
For a relativistic system
all quantities present in the Hamiltonian, in particular constraints, can involve the momentum $p_{\mu}$, positions $\stackrel[i]{}{q}^{\mu}$ and the associated momenta $\stackrel[i]{}{\pi_{\mu}}$ through Lorentz scalars. They cannot involve spacetime coordinates $x^{\mu}$ on account of translational invariance. The basic scalars out of which other scalars can be derived are $\stackrel[i]{}{q}\stackrel[j]{}{q}$, $p\stackrel[i]{}{q}$, $p\stackrel[i]{}{\pi}$, $\stackrel[i]{}{q}\stackrel[j]{}{\pi}$ and $\stackrel[i]{}{\pi}\stackrel[j]{}{\pi}$. This we assume for our class of geometric models of particles with spin. 

There ar two kinds of constraints called first class constraints and second class constraints.
     The number of arbitrary functions present in the general solution of Hamiltonian equations equals to the number of first class constraints \cite{bib:dirac}.  The first constraint, $\psi_{1}=pp-\elm^2$ commutes with all basic scalars thus also it commutes with any functions of  them. In particular, it commutes with all constraints. Hence, it is first class. The second constraint, $\psi_2=WW+\frac{1}{4}\elm^4\ell^2$, is also first class. To show this, it suffices to verify by a direct calculation that $\psi_2$ commutes
with every of basic scalars (we do not show this calculation here).  
 There will be present also a number of additional constraints.  We do not need to enter here the issue of finding constraints, we assume only that the constraints are all functions of basic  scalars. It is important that the constraints be independent.
   By a linear combination of these constraints one can bring into the first class as many constraints as possible. In effect we are left with a number $2+N_{I}$ of first class constraints $\psi_i$ (including $\psi_1$ and $\psi_2$) and an even number $N_{II}$ of second class constraints $\chi_i$, all being independent. In accordance with the Dirac generalized Hamiltonian formalism \cite{bib:dirac} for a reparametrization invariant system, all of first class constraints are used to build up the total Hamiltonian
$$H=c_1(\tau)\br{pp-\elm^2}+c_2(\tau)\br{WW+\frac{1}{4}\elm^4\ell^2}+
\sum\limits_{i=3}^{N_I+2}c_i(\tau)\,\psi_i,$$
while all of second class constraints are used to construct the Dirac bracket
\newcommand{\DB}[2]{\left\{#1,#2\right\}_{DB}}
\newcommand{\PB}[2]{\left\{#1,#2\right\}_{PB}}
$$\DB{U}{V}=\PB{U}{V}-\sum_{i=1}^{N_{II}}
\sum\limits_{j=1}^{N_{II}}\PB{U}{\chi_i}C_{ij}\PB{\chi_j}{V}, \qquad \sum\limits_{j=1}^{N_{II}}C_{ij}\PB{\chi_j}{\chi_k}=\delta_{ik}.$$
Here, $c_i$ are quite arbitrary functions of phase-space coordinates and $\PB{\cdot}{\cdot}$ stands for the Poisson bracket taken over all $4+4N_v$ position-momentum pairs, with
 $N_v$ being the number of vectors $\stackrel[i]{}{q}$ defining our particle's intrinsic structure. Let $N_c$ be the number of conditions imposed on their components. The expected number of physical degrees of freedom is thus $N_{ph}=3+4N_v-N_c$, where we have $3$ degrees of freedom for the position in space, and we must have $N_c=N_I+N_{II}/2$.
We can eliminate the arbitrary functions $c_i$ by imposing gauge conditions in the number equal to the number of first class constraints. Based on the counting of degrees of freedom according to \cite{bib:Henneaux&Teitelboim} we expect that $2N_{ph}=8(1+N_v)-2(2+N_I)-N_{II}$ since with every vector $\stackrel[i]{}{q}$ and with the spacetime-position  $x$ one associates $8$ variables in the phase space. Hence, on equating both expressions for $N_{ph}$ we come to a contradiction $1=0$, showing our expectation for the physical degrees of freedom was wrong.
We would not come to a contradiction if, instead of the two constraints $\psi_1$ and $\psi_2$, we assumed only a single condition that the Casimir mass is some function of the Casimir spin. This would result in a single first class constraint $\psi'=f(pp,WW)=0$ stating that Casimir mass and spin are functions of each other (their values will depend on the physical state).

\textit{The conclusion} is that fundamental dynamical systems (that is, those satisfying fundamental conditions) must have one less physical degrees of freedom than their phenomenological counterparts (that is, those not satisfying these conditions, see \cite{bib:astar1} for the distinction between fundamental and phenomenological dynamical systems). In the case of a fundamental dynamical system there will be a gauge freedom among the degrees of freedom originally regarded physical.

As an example consider a dynamical system described by position and the spinor direction only (then $F(\qq,\rr)=f(\qq)$ in action integral \eqref{eq:HamiltonsAction}). The expected number of physical degrees of freedom is now $3+2$ like for a rotator, thus the Hessian rank of the reduced Lagrangian should be $5$. The definition of momenta for the original Lagrangian in \eqref{eq:HamiltonsAction} lead to a  constraint $k\pi=0$ (projection invariance) for a null vector $k$, $kk=0$. These two constraints commute on the constraint surface:
$\sq{kk,k\pi}_{PB}=4\,kk\approx0$ ($\approx$ denotes the vanishing on the constraint surface). To fix the arbitrary gauge functions, we can use two gauge conditions $p\pi=0$ and $kp=\elm$, which sets the gauge corresponding to symmetries $\pi\to\pi+\lambda \,k$ and $k\to\tilde{\lambda}\, k$. Equally, well we could consider all of these four conditions together as a system of four second class constraints. Then, the first class Hamiltonian for a particle with separately fixed mass and spin  would be
$$H_r=c_t\br{pp-\elm^2}+c_{\phi}\elm^2\br{\pi\pi+\frac{1}{4}\elm^2\ell^2},$$ with quite arbitrary functions $c_t$ and $c_{\phi}$.
With the Dirac commutator
\begin{scriptsize}
$$\sq{U,V}-\br{\frac{ \sq{U,k  k}\, \sq{\pi   k,V}- \sq{U,\pi   k}\, \sq{k  k,V}  }{2} +
   \frac{ \sq{U,\pi   p}\, \sq{k  k,V}- \sq{U,k  k}\, \sq{\pi   p,V}}{2\,\elm }
   +
  \frac{ \sq{U,\pi   k}\, \sq{k  p,V}-\sq{U,k  p}\, \sq{\pi   k,V}}{\elm } }, $$\end{scriptsize}\\
 the Hamiltonian equations of motion
are (in writing them we used the constraints) $$\dot{x}=-\frac{1}{2}\,\ell^2\,\elm^3\,c_{\phi}\,k+2\,c_t\,p,\qquad \dot{p}=0,\qquad \dot{k}=2\,\elm^2\,c_{\phi}\,\pi,\qquad \dot{\pi}=\frac{1}{2}\ell^2\elm^3c_{\phi}\br{p-\elm\, k}.$$
The gauge functions can be renamed by defining new functions $t$ and $\phi$ in such a way that  $\dot{x}p=\elm\,\dot{t}$ (then $t$ is the proper time in the center of momentum frame) and $\dot{k}\pi=\frac{1}{2}\elm\,\ell\,\dot{\phi}$. Then the rapidity $\Psi$ in the center of momentum frame is ${\tanh{\Psi}}=\frac{\ell}{2}\,\abs{{\phi'(t)}}$, hence, $\Psi$ measures the angular velocity  perceived in that frame. The phase  $\int \dot{k}\pi+\dot{x}p$ behaves as $mt+\frac{1}{2}\elm\ell\phi$, however, $\phi(t)$ is left undetermined by the equations of motion. We come to a paradox, we see that rapidity $\Psi$ which one would consider a genuine physical observable shows up to be a gauge variable (!). This paradox leads us to the conclusion that the trembling motion part $-\frac{1}{2}\,\ell^2\,\elm^3\,c_{\phi}\,k$ in the velocity observable $\dot{x}$ is not physical. This part says that the particle moves with arbitrarily changing rapidity about a circle in the center of momentum frame, indeed the radius of curvature $R_c$ of the trajectory in the subspace orthogonal to $p$ is fixed by the length parameter $\ell$
$$
\begin{array}{cc}
R_c=\sqrt{-\frac{\br{\dot{x}_{\bot}\dot{x}_{\bot}}^{3}}{
\left|\begin{array}{cc}\dot{x}_{\bot}\dot{x}_{\bot}&\dot{x}_{\bot}\ddot{x}_{\bot}\\ \ddot{x}_{\bot}\dot{x}_{\bot}&\ddot{x}_{\bot}\ddot{x}_{\bot}\end{array}\right|}}=\frac{\ell}{2},
\qquad
& \begin{array}{l}\dot{x}_{\bot}=-\frac{1}{2}\ell^2\elm^3c_{\phi}\br{k-\frac{p}{m}}\\ \\
\ddot{x}_{\bot}=\br{\frac{\ud}{\ud{\tau}}{\br{\dot{x}_{\bot}}}}_{\bot}=-\ell^2\elm^5c_{\phi}^2\pi-
\frac{1}{2}\ell^2\elm^3\dot{c}_{\phi}\br{k-\frac{p}{m}}
\end{array}  \end{array}$$
and the position of the particle on that circle is undefined (arbitrary).

\medskip

Our conclusion concerning the arbitrariness in the motion of fundamental dynamical systems is true provided that 
constraints are independent. 
One cannot exclude the possibility of dependent constraints.
For example, reparametrization of a light-like velocity vector $\dot{x}$ collinear  with a null direction $k$
defining the internal structure could be absorbed by arbitrary scale of $k$.
 A kind of such particle would best resemble the Dirac electron in three respects: i) its Casimir mass and spin would be nonzero and fixed, always the same, independently of the state of the motion, ii) the intrinsic motion in the center of momentum frame would be periodic with the velocity of light, and iii) as a whole the center of mass could still move with a velocity lower than the speed of light. We leave this interesting possibility for the future.
 We note only, that \textit{formally}, one could obtain two kinds of such motion from the above Hamiltonian equations assuming two gauges: $c_t=0$ or $c_t=\elm^2\ell^2c_{\phi}$. They would correspond to two zitterbewegung states of the electron. However, such determinate motion should naturally follow from the action integral and should not be imposed by hands. The above Hamiltonian is therefore incomplete.


\end{document}